\documentclass[prc,floatfix,nofootinbib,showpacs,showkeys,preprint]{revtex4-1}
\usepackage{graphicx}
\usepackage{amssymb,amsmath,amstext,amsthm,amsfonts}
\usepackage{subfigure}
\usepackage[utf8]{inputenc}
\usepackage{float}
\usepackage{url}

\begin{document}
\today
\title{Addendum:\\ Muon-neutrino-induced charged current pion production}

\author{U. Mosel}
\email[Contact e-mail: ]{mosel@physik.uni-giessen.de}
\affiliation{Institut f\"ur Theoretische Physik, Universit\"at Giessen, Giessen, Germany}
\author{K. Gallmeister}
\affiliation{Institut f\"ur Theoretische Physik,\\ Johann Wolfgang Goethe-Universit\"at Frankfurt, Frankfurt a.\ M., Germany}

\begin{abstract}
This short paper is an addendum to a recent publication on charged current neutrino-induced pion production (Phys.\ Rev.\ C96 (2017) no.1, 015503). It presents comparisons of pion production cross sections measured at the T2K near detector for a CH target.
\end{abstract}

\maketitle
\section{Introduction}
In \cite{Mosel:2017nzk} we have discussed in some detail the seeming incompatibility of MiniBooNE and MINERvA pion production data observed by various authors. We have pointed out that new pion production data from T2K may have the potential of clarifying that 'pion-production puzzle' since T2K works with a neutrino beam with an energy distribution similar to that at MiniBooNE. We have, therefore, made a detailed comparison of GiBUU calculations with pion production data on H$_2$O in the T2K experiment as well as those from MINERvA. The main result of that study was that one and the same theory -- without any special tune -- could describe the T2K and the MINERvA data simultaneously. The present paper is an addendum to this publication with new results for a CH target at the T2K near detector (ND); final experimental analyses of data for this reaction are presently going on. The purpose of this paper is twofold: first, to help to understand the above-mentioned pion-puzzle and, second, to see if GiBUU works also for this new data set.

\section{Method}
GiBUU is a quantum-kinetic transport theory based on a non-equilibrium Green's function method \cite{gibuu,Buss:2011mx}.  It, therefore, describes only incoherent processes and does not contain any coherent production. The program version (2017) is identical to that used in \cite{Mosel:2017nzk}; again, no special tunes are used. All technical details can be found in the reference just quoted. The flux used is that of the T2K neutrino beam flux prediction 2016 \cite{T2Kflux:2016}.

\section{Results}
All cross sections are given for a CH target per nucleon; this cross section is obtained by summing the individual per-nucleon cross sections for C and H and dividing the sum by 13. The calculations use the kinematical cutoffs of the experiment shown in Table \ref{tab:cutoffs} \cite{Sanchez}.

\begin{table}[h]
\caption{Experimental phase-space limitations}
\centering
\begin{tabular}{|c|c|c|c|c|}
\hline
Observable& $\cos(\theta_\mu)> 0.2$ &$\cos(\theta_\pi)> 0.2$  & $p_\mu > 0.2$ GeV & $p_\pi > 0.2$ GeV \\
\hline
 d$\sigma/{\rm d}\cos(\theta_\mu)$ & Y & Y & Y & Y \\
\hline
 d$\sigma/{\rm d}p_\mu$& Y &  & Y &    \\
\hline
 d$\sigma/{\rm d}Q^2$& Y & Y  & Y & Y  \\
\hline
 d$\sigma/{\rm d}p_\pi$& Y & Y & Y & Y  \\
\hline
 d$\sigma/{\rm d}\theta_\pi$& Y & Y & Y & Y \\
\hline
\end{tabular}
\label{tab:cutoffs}
\end{table}

\begin{figure}[h]
	\centering
	\includegraphics[width=0.5\linewidth,angle=-90]{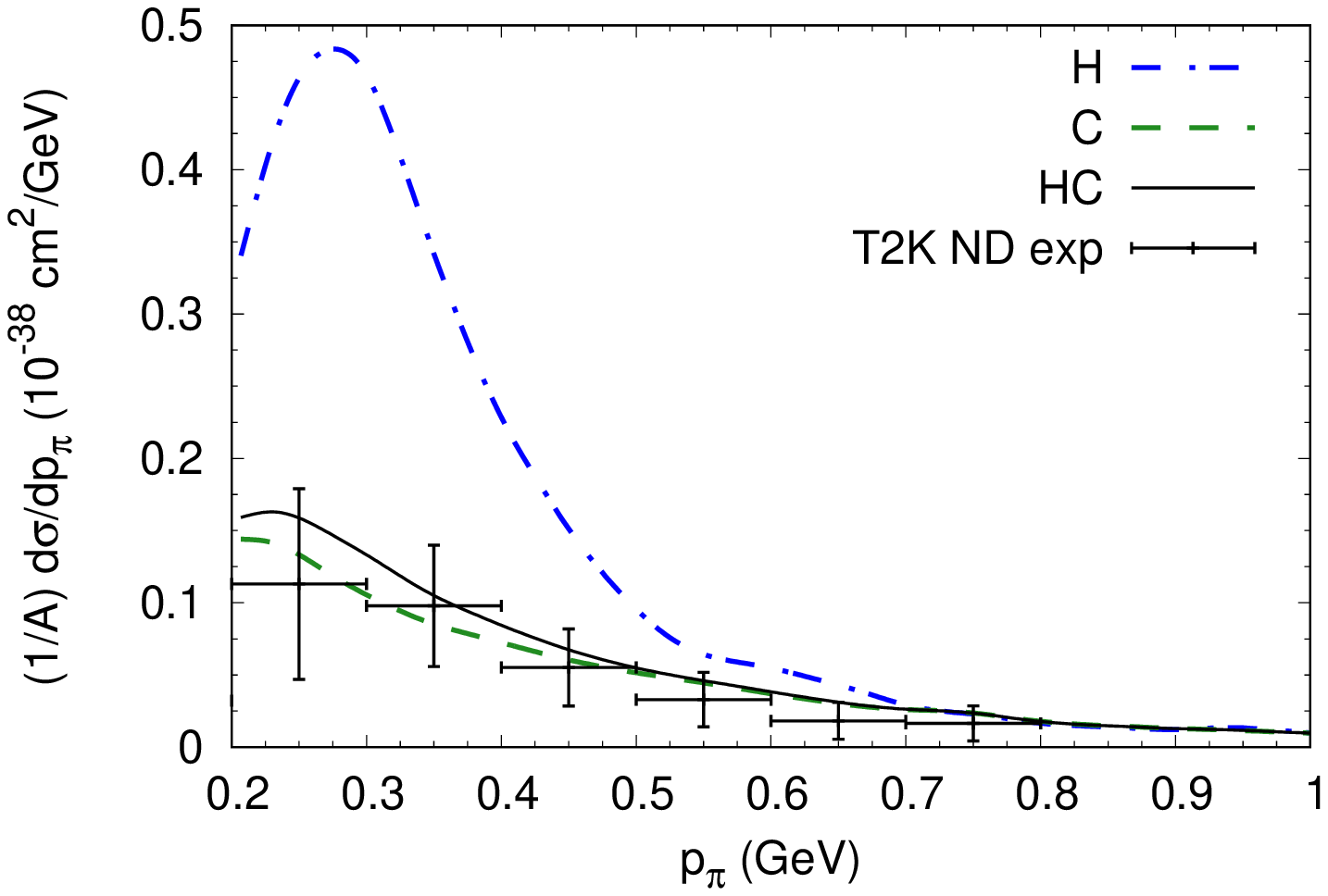}
	\caption{Pion momentum spectrum with cuts on $\pi$ and $\mu$. The uppermost (blue) dashed-dotted curve give the distribution for a H target, the (green) dashed curve that for C. The (black) solid line gives the distribution for a CH target. Data are from \cite{Castillo:2015}. }
	\label{fig:pi-spectr}
\end{figure}

Fig.\ \ref{fig:pi-spectr} shows the momentum spectrum of positively charged pions on a CH target (solid curve) together with the separate contribution from C and H. The significant difference between the C and the H spectra is a consequence of pion absorption. The total for CH is close to that for a $^{12}$C target alone.

\begin{figure}[h]
	\centering
	\includegraphics[width=0.5\linewidth,angle=-90]{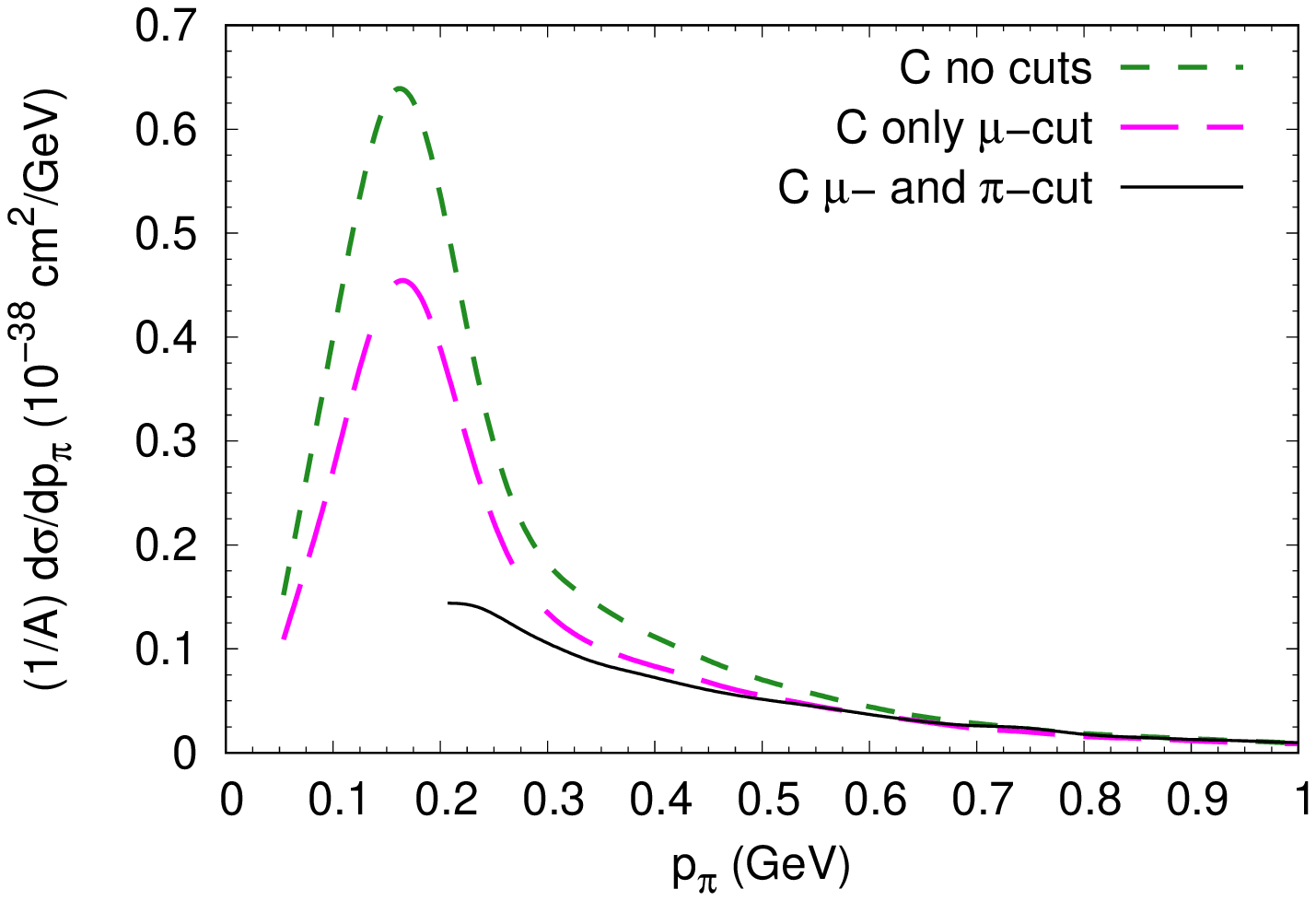}
	\caption{Effects of cuts on pion momentum spectrum for $^{12}$C.}
	\label{fig:pi-spectr-cuts}
\end{figure}
In order to illustrate the effects of the kinematical cuts in the experiment (see Table \ref{tab:cutoffs}) we show in Fig.\ \ref{fig:pi-spectr-cuts} the full pion spectrum for a C target (uppermost short dashed curve). It exhibits the well-known shape with a peak at about 0.15 GeV, just below the main absorption region. Imposing only the muon cut lowers the overall cross section, but preserves the shape (long-dashed middle curve). Cutting furthermore also on the pion, however, eliminates the peak region because the lower pion momentum cutoff is at 0.2 GeV. The experiment thus sees only a small part of the total pion production cross section.

\begin{figure}
	\centering
	\includegraphics[width=0.5\linewidth,angle=-90]{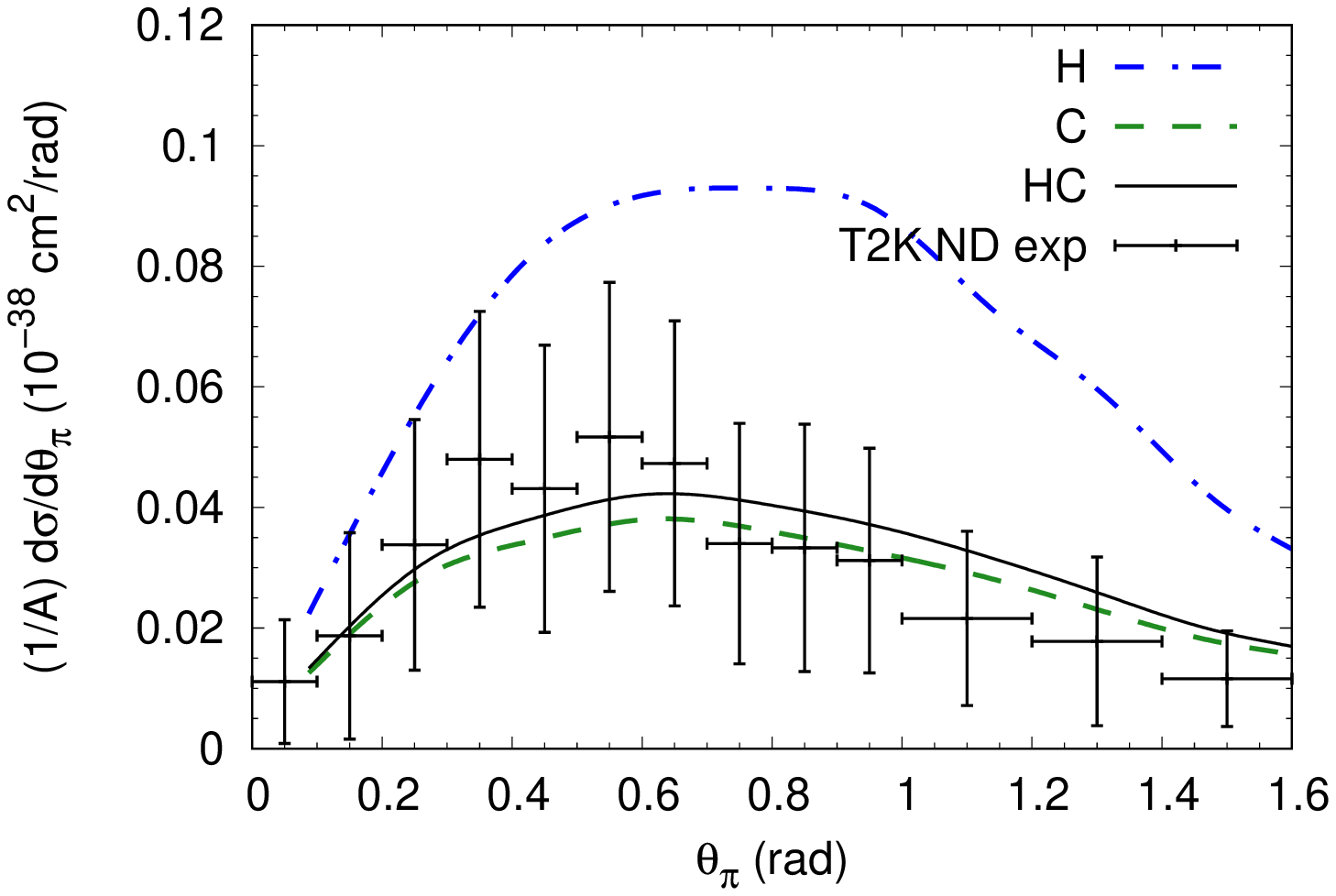}
	\caption{Pion angular distribution. Data are from \cite{Castillo:2015}.}
	\label{fig:pi-angdistr}
\end{figure}

\begin{figure}
	\centering
	\includegraphics[width=0.5\linewidth,angle=-90]{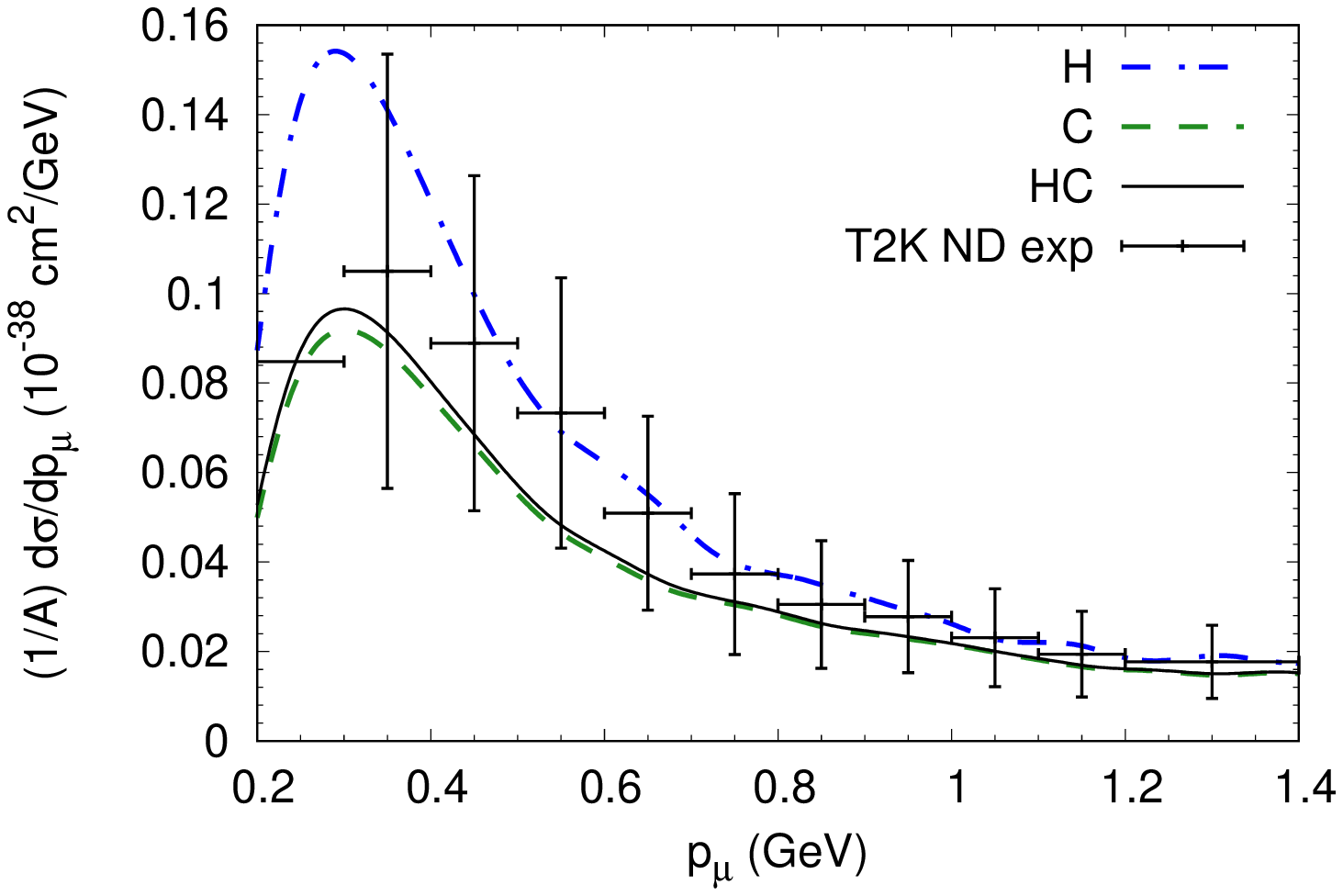}
	\caption{Momentum distribution of outgoing muons in $\pi^+$ production reactions. Data are from \cite{Castillo:2015}.}
	\label{fig:pmu-spectr}
\end{figure}

\begin{figure}
	\centering
	\includegraphics[width=0.5\linewidth,angle=-90]{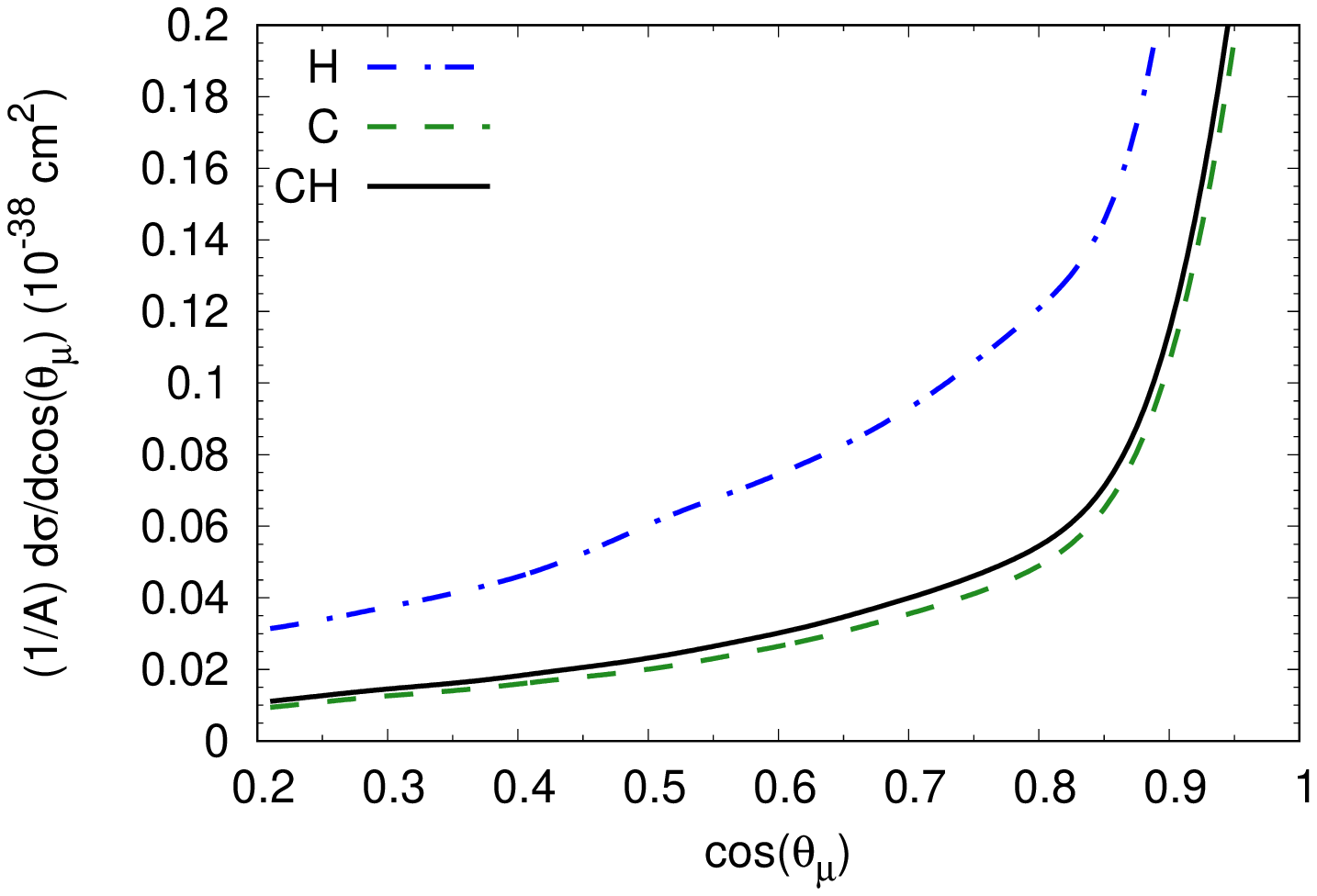}
	\caption{Angular distribution of outgoing muons in $\pi^+$ production reactions.}
	\label{fig:pmu-cosang}
\end{figure}

\begin{figure}
	\centering
	\includegraphics[width=0.5\linewidth,angle=-90]{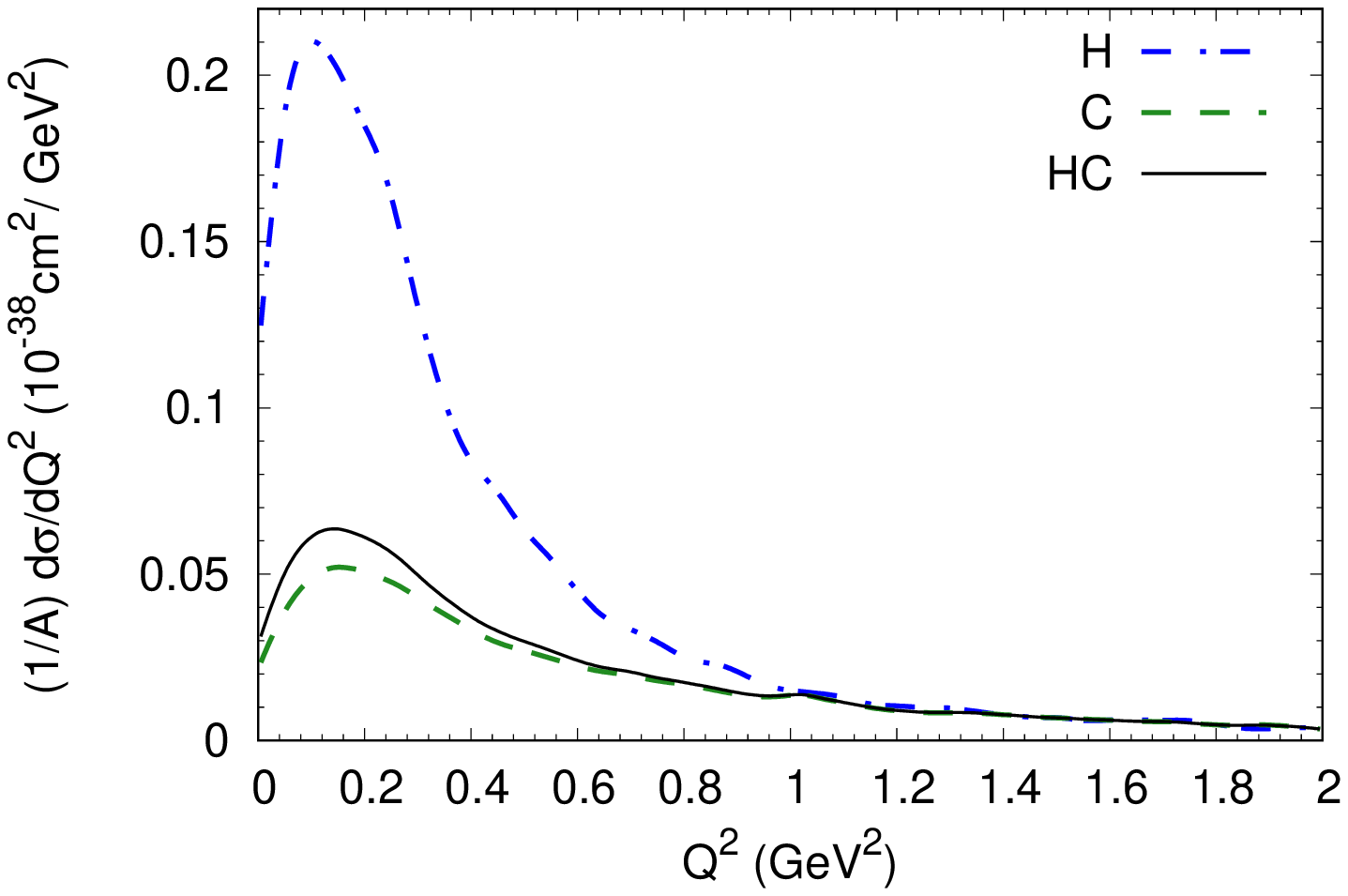}
	\caption{$Q^2$ distribution in $\pi^+$ production reactions.}
	\label{fig:Q2-distr}
\end{figure}

Figs.\ \ref{fig:pi-angdistr}, \ref{fig:pmu-spectr}, \ref{fig:pmu-cosang} and \ref{fig:Q2-distr} show the pion angular distribution, the outgoing muon's momentum and angular distribution and the $Q^2$ distribution\footnote{Data from \cite{Castillo:2015} for the $Q^2$ distribution are not shown since they have been found to be unreliable \cite{Raquel:2018}}.

As mentioned earlier, GiBUU produces only cross sections for incoherent processes. Any coherent production is expected to add some strength at small $Q^2 < 0.2$ GeV$^2$. It is interesting to note that the pion angular distribution in Fig.\ \ref{fig:pi-angdistr} is described quite well also for forward angles indicating only a small coherent contribution. In contrast, the comparison with MINERvA data in Fig.\ 5 of Ref.\ \cite{Mosel:2017nzk} showed a disagreement at forward angles $\lessapprox 40^\circ$, i.e.\ $\approx 0.7$ rad. This could be due to contributions of coherent pion production which is known to increase with neutrino energy \cite{Martins:2016vts}. On the other hand, the MINERvA analysis \cite{Eberly:2014mra} used cuts not only on the incoming flux, but also on the invariant mass ($W < 1.4$ GeV). The $W$ here was that appropriate for an incoming four-momentum $Q^2$ and energy transfer $\nu$ on a nucleon at rest ($W^2 = M^2 + 2M\nu -Q^2$). Together with the incoming neutrino energy also $Q^2$ and $\nu$ had to be reconstructed thus introducing some generator dependence into the pion production data of \cite{Eberly:2014mra}.

\begin{figure}
	\centering
	\includegraphics[width=0.5\linewidth,angle=-90]{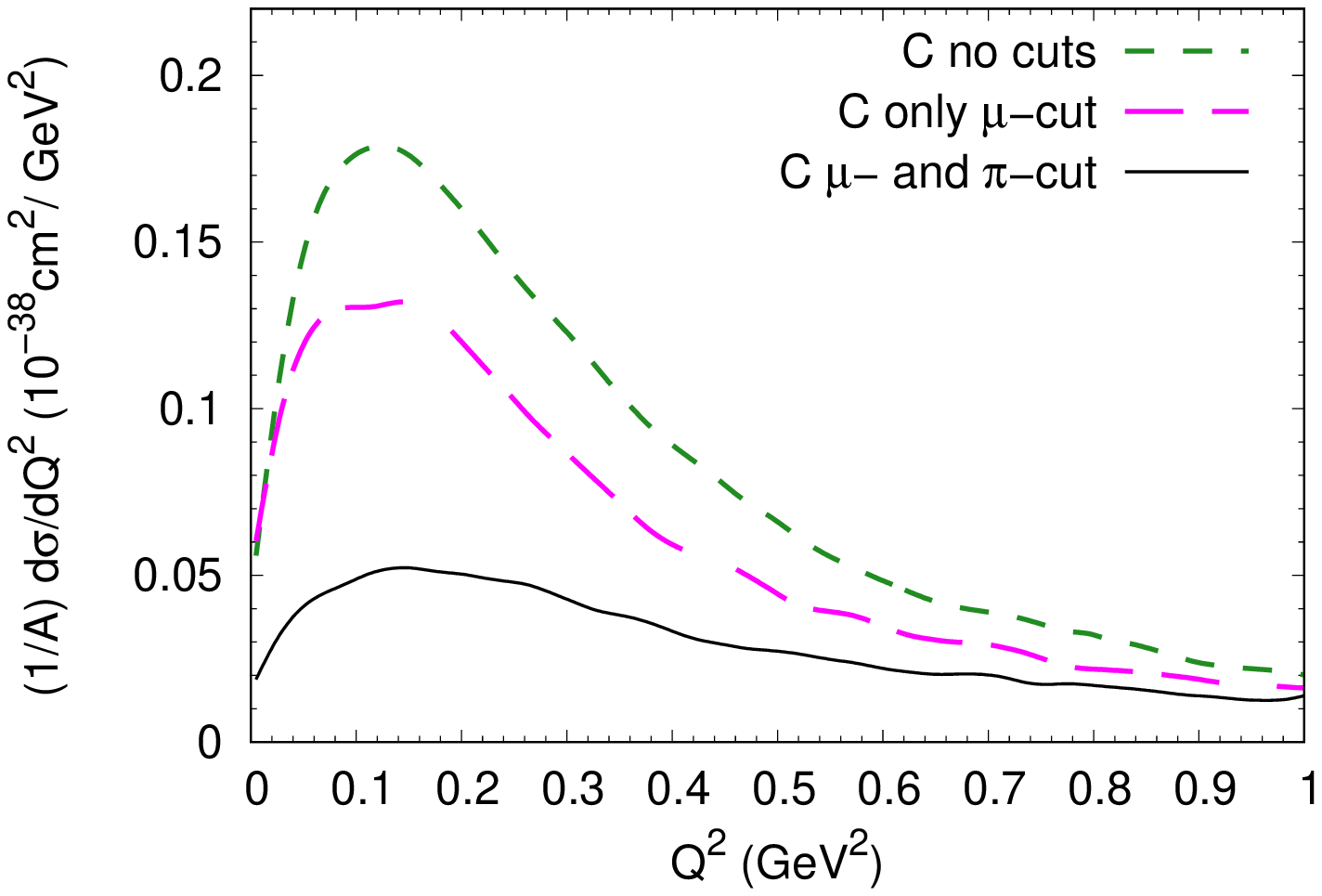}
	\caption{Effects of cuts on $Q^2$ distribution for $^{12}$C target in $\pi^+$ production reactions.}
	\label{fig:Q2-distr-cuts}	
\end{figure}

In Fig.\ \ref{fig:Q2-distr-cuts} we illustrate the effects of the acceptance cuts on the $Q^2$ distribution. The sizeable difference between the uppermost short-dashed curve (without any cuts) and the lowest solid curve again illustrates that the experimental acceptance limitations cut out a large part of the cross section.

\section{Conclusions}
Understanding pion production in neutrino-nucleus reactions is essential since pion production makes up for about 1/2 - 2/3 of the total cross section at DUNE. While present theories cannot describe the MiniBooNE data (with a disagreement both in shape and magnitude of the cross section), we have shown in \cite{Mosel:2017nzk} that GiBUU describes both the MINERvA pion production data on CH as well as the T2K data on H$_2$O quite well, without any special tune. In this paper we have now shown that the same degree of agreement of GiBUU calculations with pion production data is also reached for the T2K flux on CH. It will be interesting to see if the additional T2K data, presently still under analysis, also agree with these predictions.

For a comprehensive test of our understanding of neutrino-pion production on nuclei experiments  are needed that have a significantly larger phase-space coverage than in the T2K ND.

For further tests it is also essential to obtain pion production data for the heavier target $^{40}$Ar, the detector material in DUNE, at the higher energies expected in the DUNE flux. The recent data from ArgoNeuT are promising; again good agreement of GiBUU calculations with the data is obtained \cite{Acciarri:2018ahy} while other generators all come out too high. We speculate that the latter disagreement could be due to the decoupling of pion production and pion absorption in these other generators.  Pion production and absorption through any resonance are linked by time-reversal invariance of the $\pi N \leftrightarrow N^*$ interaction. In contrast to other generators, GiBUU respects this connection.

\begin{acknowledgments}
We acknowledge helpful information from F.\ Sanchez and R. Castillo Fernandez on the acceptance limitations of the T2K experiment. One of the authors (U.M.) is also grateful to R. Castillo Fernandez for discussions on the data evaluation.
\end{acknowledgments}

\end{document}